\documentstyle[11pt,paspconf]{article}

\begin{document}

\title{Biases in the Main Sequence Fitting Distances to Globular Clusters 
based on the Hipparcos Catalogue}

\author{Eugenio Carretta and Raffaele G. Gratton}
\affil{Osservatorio Astronomico di Padova, vicolo Osservatorio 5, I-35122, 
Padova, ITALY}

\author{Gisella Clementini and Flavio Fusi Pecci\altaffilmark{1}}
\affil{Osservatorio Astronomico di Bologna, via Zamboni 33, I-40126 Bologna, 
ITALY}

\altaffiltext{1}{Stazione Astronomica di Carloforte, I-09012, Capoterra, 
Cagliari, ITALY}

\begin{abstract}
We discuss the different biases affecting the sample of field subdwarfs
selected from the Hipparcos Catalogue, and used in the Main Sequence Fitting
technique to derive distances to Galactic Globular Clusters.
The adopted average corrections significantly affect the
derived distance moduli, explaining the differences among various
groups using this technique.
\end{abstract}



\section{How many biases? How large is their effect?}

Several biases affect the derivation of distances to globular clusters (GCs) 
$via$ Main Sequence Fitting method. 
However, their impact can be rather different, depending on how the field
subdwarfs used to build up the template sequences are selected.

In a {\it a priori} selected sample, selection criteria do not rest on 
parallaxes, but 
on other constraints such as
the metal abundance of the subdwarf sample (Gratton et al. 1997, G97; Reid 1997,
1998, R97, R98; Carretta et al. 1998, C98; a subset of the sample in Pont et
al. 1998, P98).
It is usually easy to correct for biases, in this case, since selection 
criteria
are rather well defined by the investigators themselves.

In a {\it a posteriori} selected sample, objects are extracted from the
colour-magnitude diagram of the whole HIPPARCOS
catalogue, for instance selecting stars within a given range of M$_V$ (i.e. 
parallax) and colour (i.e. metallicity, with metal abundances 
derived from colours; the second subset of stars used by  P98).
Selection criteria are rather poorly defined in this case, and it is much more 
difficult to correct for biases.
\begin{table}
\caption{Systematic effects to be considered} 
\begin{center}\scriptsize
\begin{tabular}{lrr}
Effect & & $\Delta (m-M)$  (C98)\\
\tableline
 Malmquist bias  &  $\Rightarrow$ &  negligible\\
 Lutz-Kelker  &  $\Rightarrow$ &  $\pm 0.02$\\
 Metallicity bias  &  $\Rightarrow$ &  depends on the sample\\
 Binaries (in the field) &  $\Rightarrow$ &  $\pm 0.02$\\
 Binaries (in clusters) &  $\Rightarrow$ &  $\pm 0.03$\\
 Non solar abundance ratios  &  $\Rightarrow$ &  negligible\\
 Photometric calibrations &$\Rightarrow$ &$\pm 0.04$ (mean value from 9 GCs)\\
 Reddening scale  &  $\Rightarrow$ &  $\pm 0.07$\\
 Metallicity scale  &  $\Rightarrow$ &  $\pm 0.08$\\
Total uncertainty ($1\sigma$)&  $\Rightarrow$ &  $\pm 0.12$ \\
&&(at least...)
\end{tabular}
\end{center}
\end{table}

Major systematic effects affecting the Main Sequence Fitting technique are 
listed in Table 1 and briefly reviewed below.

{\bf (a) Malmquist bias -- } 
In magnitude limited samples (with a range in luminosity) observational 
errors produce a bias toward 
luminous objects. In turn, the
average absolute magnitude  is systematically overestimated.
Corrections are negligible in G97 and C98 where the distribution in M$_V$
is narrow (only unevolved main sequence stars were used). The effect, 
however, must be considered in R97 and P98,  
whose samples include also stars brighter than the turn-off.

{\bf (b) Lutz-Kelker effect --}
If stars are selected using parallax as a criterion and/or are weighted
in the sample according to the ratio $\sigma_\pi/\pi$, average parallaxes
are overestimated.
Corrections strongly depend on $\sigma_\pi/\pi$, and 
on the parallax distribution of the original population,
which is not known, but can be estimated from the proper motion
distribution (see Hanson 1979), as done in R97. 

Care should be taken when Lutz-Kelker corrections are large. A Montecarlo 
approach should be preferred in these cases (G97, P98).
However, the safest procedure is to use only stars with good
parallaxes ($\sigma_\pi/\pi<0.12$), since corrections are small 
($<0.02$~mag for the weighted average of the sample in C98).

{\bf (c) Metallicity bias --}
Since the metallicity distribution of the stars in the solar neighbourhood is 
strongly skewed toward solar values, when colours are used to extract a sample 
of 
metal-poor stars from the Hipparcos catalogue, random errors will 
increase the number of metal-rich stars erroneously measured
too blue with respect to the number of metal-poor stars measured
too red. 
This is further complicated due the non-gaussian distribution of the errors 
in the colours taken from Tycho catalogue. 
This metallicity bias affects only samples derived {\it a posteriori} 
(the second sample in P98).

{\bf (d) Binaries (in the field) -- }
A too long distance scale will be derived if the 
local subdwarf sample is contaminated by unresolved binaries, since the
combined systems would be redder and brighter. 
The safest approach is to use only {\it bona fide} single stars 
(G97, C98). However, corrections may be required, since some residual 
undetected binary  may escape detection.
It is not easy to derive accurate binary corrections, since they depend on
rather uncertain parameters:
(i) the actual incidence of binaries and (ii) the distribution in mass 
(or luminosity) of the secondary components. 

P98 assumed that half of the stars are binaries, and that the correction for 
each binary is, on average, half the maximum value. The net result was a
rather large estimate of  0.18 mag for this correction, applied to known and 
suspected binaries, as well as to {\it bona fide} single stars.

G97 compared binaries and {\it bona fide} single stars: from the average offset
of the colours and the scatter around it they derived an average
correction for each binary of $0.16\pm 0.05$~mag, and a fraction $p\leq 0.16$\
of undetected binaries in their subdwarf sample (beside the 41\% of
stars which are known or suspected binaries). The average binary correction
derived by G97 (to be applied $only$ to {\it bona fide} single stars)
is of $0.02\pm 0.01$~mag.

{\bf (e) Binaries (in globular clusters) --}
If contamination by binaries is present in the main sequence
of a GC, a too short distance scale will be derived.

Binary incidence is likely to be higher in the field, where
multiple systems have higher probability to survive disruption. However,
blending of unrelated stars due to the extreme crowding conditions 
typical of GCs, has the
same photometric effect of physical binarity.
To reduce this problem, the usual approach is to use modal rather than mean 
values to identify the main sequence mean loci of globular clusters.

{\bf (f) Photometric calibrations --}
The main sequence is a rather  steep relation between colour and magnitude
[$4<dV/d(B-V)<7$ for unevolved dwarfs]. Therefore, very accurate photometric
data are required, since any error in the intrinsic colours of either
sequences translates into (large) errors in the derived  magnitudes. 

In principle field and cluster stars should be observed with the 
same instrumental set up and on the same
photometric system. But the much brighter field stars are usually observed 
with photomultipliers, while data for GCs are obtained with CCDs, and 
their magnitudes transformed to the chosen system by means of standard stars. 
Uncertainty in the derived transformations may result in quite large
errors (from 0.02 up to 0.04 mag) for individual clusters. 
For instance, an offset of 0.04 mag exists in the MS colour of M92
(the only cluster considered in P98)
between  Heasley \& Christian (1991) and Stetson \& Harris 
(1988). This by itself implies  an  uncertainty of 
about 3-4 Gyr in the age derived for this cluster. A safer approach 
is to average results 
over a large number of clusters (see R97, R98, G97, C98).

{\bf (g) Reddening scale --}
A uniform scale should be used for field subdwarfs and GC stars. 
Two reddening scales have been used for the subdwarfs:
R97, G97 and C98 used
reddening estimates from Carney et al. (1994), Schuster \& Nissen (1989), and
Ryan \& Norris (1991), while reddenings used by P98 (from Arenou et
al. 1992) are on average 0.016 mag larger, resulting into a shorter distance 
scale.
For the GC reddenings, however, all authors used the same scale (Zinn 1980).

If one compares adopted data with cosecant-laws for reddening, GCs and
subdwarfs result on a uniform scale if the scale-height of the 
galactic
dust disk is 100~pc, for the reddenings adopted by G97, C98 and R97, and 40 pc
if the values by  P98  are used. Since the former value is in good agreement 
with current determinations of the galactic dust scale-height, 
while the latter is at the lower extreme of the admitted range, the reddening scale for subdwarfs has to be considered still quite 
uncertain ($\pm 0.015$ mag).

{\bf (h) Metallicity scale --}
A systematic difference of 0.1 dex in the metallicity scale adopted for
subdwarfs and GCs translates into an error of $\sim 0.07$~mag (0.03~mag at
[Fe/H]=$-2$, and 0.11 mag at [Fe/H]=$-$1.0, see G97 and C98 for details).

A uniform abundance analysis has already been used by G97, 
R98 and C98; however, since abundances for GCs are derived from giants 
rather than dwarfs, some differential
effect ($\sim 0.1$~dex) cannot be excluded.
High dispersion analysis of
dwarfs and turn-off stars in GCs would be crucial, but this 
test must wait for high resolution spectrographs mounted at 8~m class 
telescopes (UVES at VLT).

{\bf (i) Non solar abundance ratios --}
The overabundance of $\alpha-$elements found 
in metal-poor stars means that extra electron-donors  
(hence, additional electron pressure) are available, as well as a larger 
blanketing. 
Clementini et al. (1998)
made some estimates scaling the abundances in the 
stellar model atmospheres, usually solar scaled, for this 
overabundance of electron donors (i.e. the metallicity of the model 
atmosphere was scaled down as [(Mg+Si+Fe)/H]). On the Fe I lines usually 
adopted for abundance analysis the effect turned out to be quite small 
($\sim 0.02$ dex). 

\section{Summary and conclusions}

The advent of the Hipparcos parallaxes has allowed to strongly improve the GCs
Main Sequence Fitting Technique:
the major contribution to the total error budget has moved from parallaxes
to  photometric calibrations, reddening and metallicity scale.
But, despite the invaluable contribution of the Hipparcos results, the Main
Sequence Distances to Galactic Globular Clusters still suffer from  a
total uncertainty of $\pm 0.12$ mag at least. Therefore, presently, neither
the short nor the long distance scale can be completely ruled out.


\begin{references}
\reference Arenou, F., Grenon, M., \& Gomez, A. 1992, A\&A 258, 104
\reference Carney, B.W., Latham, D.W., Laird, J.B., \& Aguilar, L.A. 1994, AJ,
 107, 2240
\reference Carretta, E., Gratton, R.G., Clementini, G., Fusi Pecci, F. 1998 
 in preparation (C98)
\reference Clementini, G., Gratton. R.G., Carretta, E., Sneden, C. 1998, MNRAS, 
 in press
\reference Gratton, R.G., Fusi Pecci, F., Carretta, E., Clementini, G., 
 Corsi, C.E., Lattanzi, M.G. 1997, ApJ, 491, 749 (G97) 
\reference Hanson, R.B. 1979, MNRAS, 186, 675
\reference Heasley, J. N., \&  Christian, C. A. 1991, AJ, 101, 967 
\reference Pont, F., Mayor, M., Turon, C., Vandenberg, D.A. 1998, 
  A\&A, 329, 87 (P98)
\reference Reid, I.N. 1997, AJ, 114, 161 (R97)
\reference Reid, I.N. 1998, AJ, 115, 204 (R98)
\reference Ryan, S.G., \& Norris, J.E. 1991, AJ, 101, 1835
\reference Schuster, W.J., \& Nissen, P.E. 1989, A\&A, 221, 65
\reference Stetson, P. B., \&  Harris, W. E. 1988, AJ, 96, 909 
\reference Zinn, R. 1980, ApJS, 42, 19 
\end{references}
\end{document}